\documentclass[12pt]{iopart} 
\pdfoutput=1
\usepackage{graphicx} 
\usepackage[caption=false]{subfig}
\usepackage{color}
\usepackage{float}
  
\graphicspath{{./Figures/}}    
  
\begin{document}

\title[]{Collective movement 
in alarmed animals groups: 
a simple model with 
positional forces and a limited attention field} 

\author{A. M. Calv\~ao$^{\ddag}$ and E. Brigatti$^{\star}$}

\address{$^{\ddag}$ Instituto Polit\'ecnico do Rio de Janeiro, Universidade do Estado do Rio de Janeiro,  Rua Bonfim, 25, Vila Amélia, 28625-570, Nova Friburgo, RJ, Brazil}
\address{$^{\star}$ Instituto de F\'{\i}sica, Universidade Federal do Rio de Janeiro, 
Av. Athos da Silveira Ramos, 149,
Cidade Universitária, 21941-972, Rio de Janeiro, RJ, Brazil}
\address{e-mail address: edgardo@if.ufrj.br}

\maketitle
  
\begin{abstract}

We perform a numerical analysis of a recent introduced model for describing 
collective movement in alarmed animals groups.
This model, derived from a position-based interaction and a limited attention field,
displays a non-equilibrium phase transition between a crystal-like phase,
characterised by a single densely packed cluster of individuals, 
which occurs for small attention field angles,
and a disordered phase, constituted of  fragmented clusters,
which emerges at large angles. 
The transition is quantitatively characterised,  measuring its critical point 
and describing its scaling behaviour, which 
suggests that the nature of the transition is discontinuous.
Finally, we study the correlations in the directions of motion, which 
show strong positive values at short distances. This fact suggests
that individuals' effective perception range goes far beyond the 
model interaction range. 

\end{abstract}



\section{Introduction}



A particularly interesting aspect of collective animal motion is the spontaneous and abrupt aggregation  manifested by alarmed group of social animals \cite{hamilton,viscido1,morton,devos}.
This problem gained a renewed interest  with some recent 
works on actual herds \cite{toulet}. Such field studies undertook quantitative analysis under controlled experimental conditions. The work of Ginelli {\it et al.}  \cite{ginelli15} shown an intermittent behaviour between two conflicting tasks: foraging,
realised by dispersed individuals, 
and protection, offered by aggregation in a compact group. 
This second aspect was previously studied by
detecting sheep movements  in response to the approach of a dog in a 
field setting \cite{king}.
Trajectories were obtained by using a GPS technology 
and shown that animals manifest a strong attraction towards the centre of 
the threaten flock.
As described by modelling and theory \cite{hamilton,krause}, 
as a consequence of a possible predation,
animals aggregate by means of a fast collective movement. 
A classical hypothesis explaining this 
gregarious 
behaviour is the so called ``selfish herd hypothesis", which suggests that cohesive swarms are generated because each member of the group moves toward neighbours for minimising its own predation risk \cite{hamilton}.
Unfortunately, the implementation of this reasonable idea into a specific and practical 
algorithm is not so straightforward.
The original Hamilton's algorithm \cite{hamilton}, where the focal individual moves towards the nearest neighbour, was a first tentative in this direction but it failed in describing
compact and dense groups. In fact, in two dimensions, fragmentation into 
multiple sub-groups of a few individuals occurs \cite{morton, viscido2}.
For this reason, discovering an elementary movement rule which can produce compact
aggregation is still an unsolved problem, 
denominated the ``dilemma of the selfish herd''.

This movement rule must satisfy some 
natural constraints, and the solution of the problem
must be reached constructing a model built up on perceptive bases.
This means that individuals movements should be based  exclusively 
on the instantaneous visual information they obtain and 
process.
Documented physiological and cognitive facts must be taken into account.
For example, some type of selection of the interaction partners must be introduced.
In fact, it has been shown that animals generally interact with a small number
of neighbours \cite{cavagna,ballerini,mosquitofish}
and a limited attention field is a general well-known characteristic 
of cognitive processes. 
An extreme 
example of how much visual attention is selective
is the 
phenomenon called ``inattentional blindness'',
which suggests that humans perceive 
only those objects and events that receive focused attention \cite{simons}.

Moreover, the 
model must be assembled from basic 
mimetic rules which operate just for the 
punctual
duration of the stimulus.
The considered  individual movement rule must be as simple as possible.
This issue is particularly important because the rule must satisfy two biological considerations.
The first is that even very simple animals, with limited cognitive ability, like arthropods,
must be able to follow it.
The second is that many different species should be able to act in accordance with it, 
from Crustacea \cite{viscido1} to Mammals \cite{devos}.
Finally,  a behavioural ecology perspective, 
suggests that the rule must be 
simple enough for allowing 
natural selection 
to favour and fix it \cite{reluga}.
Usually considered interactions, as for example molecular type forces, 
with a weighted sum of a large number of 
neighbours, clearly do not satisfy these assumptions.
Not even models based on local velocity, which imply
a memory based mechanism, are coherent with these considerations.

A possible solution to this problem  was recently introduced in \cite{edgardo},
with the adoption of a new algorithm. 
This algorithm added a simple 
ingredient to the Hamilton's rule: 
the fact that 
animals can exercise their attention only over a limited horizon \cite{lemasson}.
The analysis of this model 
demonstrated that, if animals try to approximate the 
first neighbour enclosed in a specific attention field, it is possible to 
produce the densest aggregation in a centrally compact swarm \cite{edgardo}.
These results are obtained for selected values of the angle which 
defines the attention field and the compact swarms are robust to the introduction
of noise.
The general results produced by this
model can be connected with some 
reported behaviours of animals in the wild. 
In \cite{edgardo},
a qualitative comparison of the model 
output with some 
empirical observations of  groups of crabs  \cite{viscido1} was presented.
These results were obtained calibrating the model 
parameters with realistic values measured in \cite{viscido1}. 
For the opportune values of the attention field
lively shrunk configurations
qualitatively very similar to the observed real data
were obtained.
Moreover, for this set of parameters,
the typical time necessary to reach the compact configurations 
resulted to be comparable with the experimental one. 
In fact, the calibration of the animals velocity 
implied fixing the unit of the simulation time step.
It followed that a significant reduction
of the domain of danger was obtained after a few seconds,
a result that can be realistically compared with 
field data \cite{edgardo},
validating the model as capable of describing 
an actual process in ecology.
\\

It is important to note that this model of collective movement
is based on an attraction-repulsion interaction 
which depends on the instantaneous position of neighbouring individuals
and not on their velocity.
The simple introduction of an attention cone is sufficient to generate a very rich behaviour.
These two fundamental ingredients were introduced for the first time for solving the ``dilemma of the selfish herd'' in \cite{edgardo}, and they have recently gained a lot of attention in the
literature.
Different works investigated the impact of the introduction of a 
vision cone in other type of models \cite{cone}, and, more recently,
a positional-based force with a vision cone have been applied
in the classical framework of flocking models without cohesion 
in a 
study written by Barberis {\it et al.} (see \cite{fernando}).\\




The principal purpose of this work is to quantitatively explore  
the model first introduced in \cite{edgardo} throughout an accurate numerical analysis.
First, we outline the general model dynamical behaviour.
Second, we characterise the process of compact
aggregation 
in term of a phase transition. 
Finally, we studied the properties of the correlations 
in the direction of motion of the individuals.


\section{The model}

We build up a model which considers only the movements 
of the individuals relative to the centre of mass of the group.
This means that the contribution of the global translation and rotation 
of the swarm is subtracted from the individuals movements. 

\begin{figure}
\begin{center}
\includegraphics[width=0.8\textwidth,angle=0]{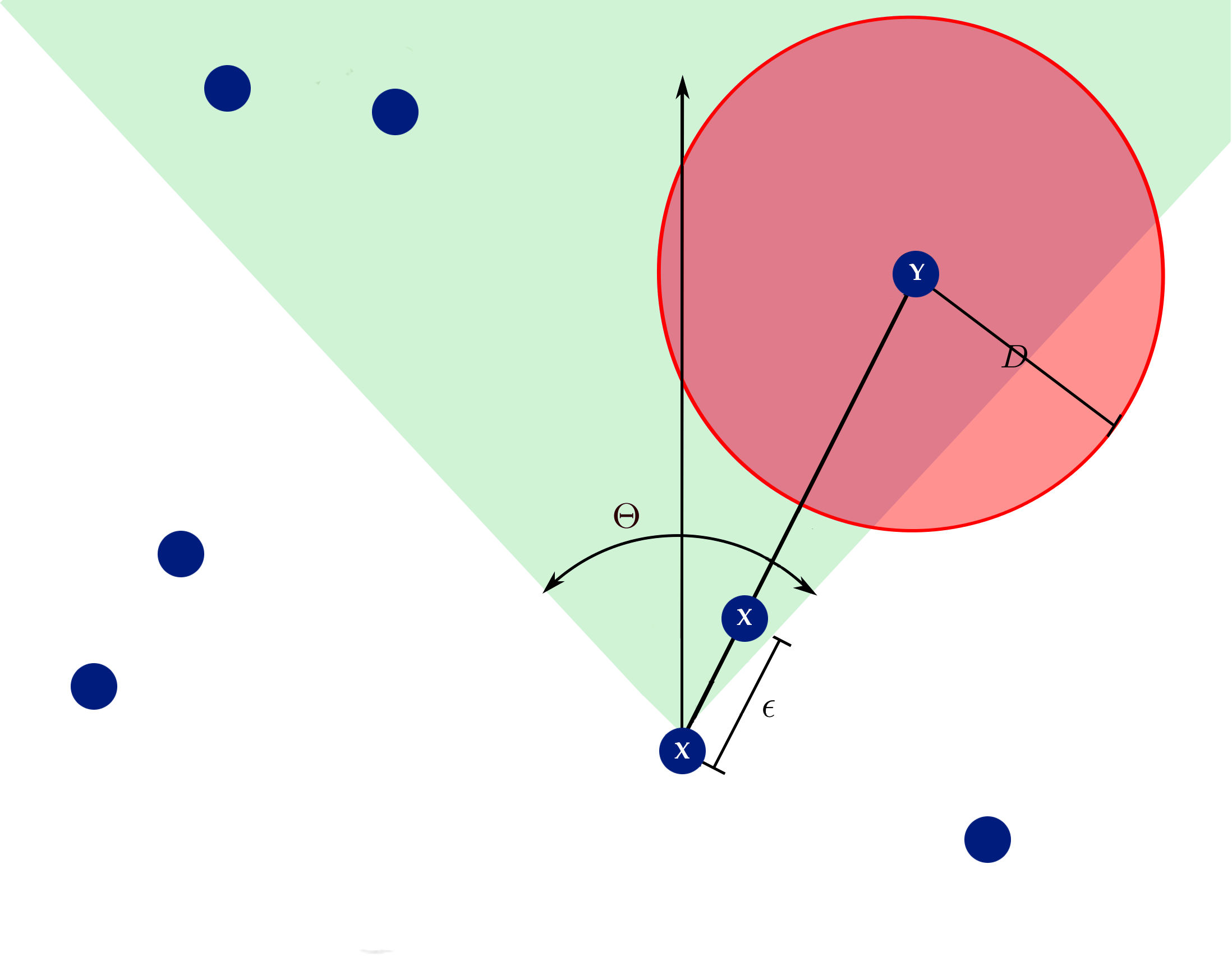}
\caption{\small {Agents are represented by the blue circles. 
Agent $X$ 
 is randomly selected and a 
 gazing direction (bisector of the angle $\Theta$) is assigned.
The attention field, where the nearest neighbour is searched (agent $Y$)
is the green region. 
The stress zone of individual $Y$ is the red disc. 
}
\label{fig:steps}}
\end{center}
\end{figure}

The basic rule which governs our model is very simple: every agents 
seek for an optimal distance from the nearest 
topological neighbour encompassed in 
the attention field.
This is implemented by means of the following algorithm:
an agent $X$ and its gazing direction are randomly selected.
The gazing direction is the bisector of the angle $\Theta$, 
which defines the cone (attention field) where neighbours are perceived.
The nearest topological neighbour $Y$ contained in the attention field  
is identified.
Agent $X$ gives a step $\epsilon$ towards $Y$, 
trying to reach an optimal distance $D$.
In detail, 
considering $x_t$ the position of agent $X$ and $y_t$ the one of $Y$, 
if $|{\bf x_t}-{\bf y}|>D$, $X$ moves in the direction 
of $Y$ a step $\epsilon$: 
${\bf x_{t+1}}={\bf x_t}+\epsilon {\bf v}$, where
${\bf v}$ is a unit vector.
If this movement causes the stress zone invasion, 
$X$ stops at a distance $D$ from $Y$.
If $|{\bf x_t}-{\bf y}|<D$, the motion is: 
${\bf x_{t+1}}={\bf x_t}-\epsilon {\bf v}$.
If this motion leaves $X$ farther than $D$ from its neighbour,
$Y$ stops at a distance $D$.
Movements are determined asynchronously and the
time step $t$ is incremented after the updating 
of all the agents' positions.
Figure \ref{fig:steps} shows an example of these rules. 

The system is composed by $P$ agents which move continuously
on a square of linear size L and which do not cross the given boundaries.
As initial condition, individuals are uniformly and randomly distributed on the square.
In the following, we set, without loss of generality, $D = 1$ and 
$\epsilon  = 0.1$, and we express all 
scales in terms of these units.
We run different simulations varying the number of individuals but 
maintaining fixed the density ($\delta=0.3$). 
This was obtained modifying the square size ($L=\sqrt{P/0.3}$).
This is a standard approach in finite size scaling analysis, 
as the model behaviour depends on the density, as suggested in \cite{edgardo}. 

\section{Results and discussion}

The dynamics behaviour of the model is described
introducing three different order parameters
which can characterise 
the degree of order and cohesion of the system.

The degree of positional orientational order 
of a configuration is obtained introducing,
 for each organism  \cite{strandburg}:
$\psi_j= \frac{1}{N_j}\sum^{N_j}_{k=1} \exp(i6\phi_{jk}),$
where $N_j$ is the number of topological neighbours of 
individual $j$, 
which are the animals whose Voronoi polygons share an edge with $j$.
The neighbours' index  is $k$ and 
$\phi_{jk}$ is the angle relative to the bond between 
$j$ and $k$ and an arbitrary fixed reference axis.
A positional orientational order parameter is estimated
calculating the norm 
of $\psi_j$ averaged
over all individuals $j$:
$\psi_6= \frac{1}{P}\left| \sum^{P}_{j=1} \psi_j \right|.$

Translational order is described counting
the number of individuals 
contained in a circle of radius $D+0.001$ centred on a given animal.
The translational ordering parameter $N$ is obtained 
averaging this quantity over all individuals.

Finally, the level of swarm cohesion is measured 
evaluating the quantity: 
\begin{equation}
C(t)= \frac{a(t)}{a(0)}
\label{eq:C}
\end{equation}

where $a(t)$ is the 
convex hull of the finite points set corresponding to the given group
at time $t$.
The convex hull of a set of points is the area of the smallest 
convex polygon encompassing all the points and 
it is a good approximation of the 
area covered by the swarm at a given time $t$.
It has been already used for characterising group
cohesion in previous works \cite{edgardo, ginelli15}.
$C(t)$  estimates what in the classical ecological literature is called 
''the reduction of the overall domains of danger'' \cite{hamilton}.
The domain of danger indicates the area closer to a given individual 
than to any other group member and it is estimated considering the Voronoi polygon 
constructed around the group member \cite{hamilton}. 
The overall domains of danger is 
obtained summing all the Voronoi polygons, excepted 
the ones corresponding to the edge individuals \cite{viscido1}. 
A more efficient evaluation of  
this quantity is obtained measuring the convex hull of the group. 



\begin{figure}[h]
\begin{center}
\includegraphics[width=0.75\textwidth, angle=0]{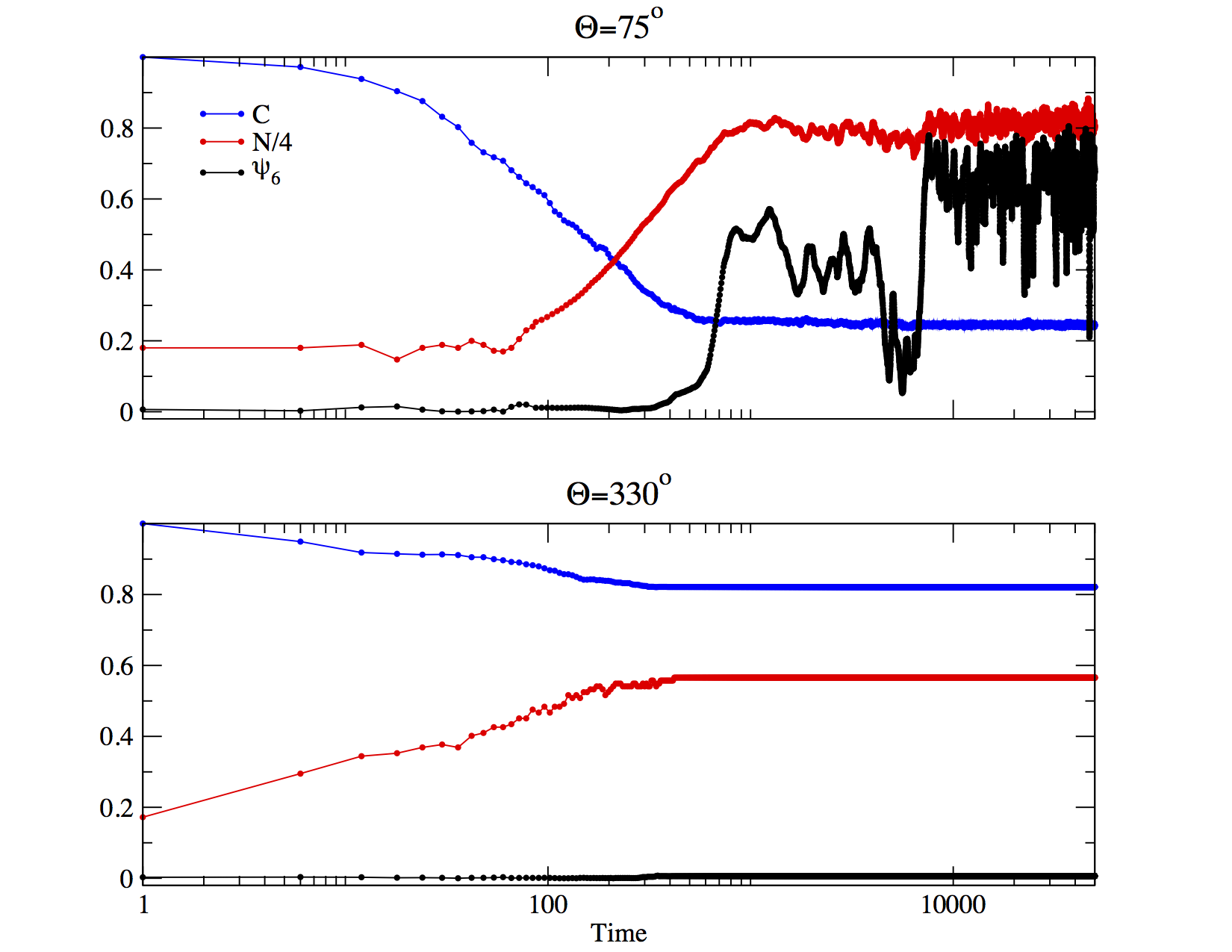}
\end{center}
\caption{\small {Time evolution of the order parameters $N$, $\psi_6$
and $C$ for $\Theta=75^{\circ}$ and $\Theta=330^{\circ}$;
$P=61$.
}
}
\label{fig_Dynamics}
\end{figure}
\begin{figure}[h]
\begin{center}
\includegraphics[width=0.55\textwidth, angle=0]{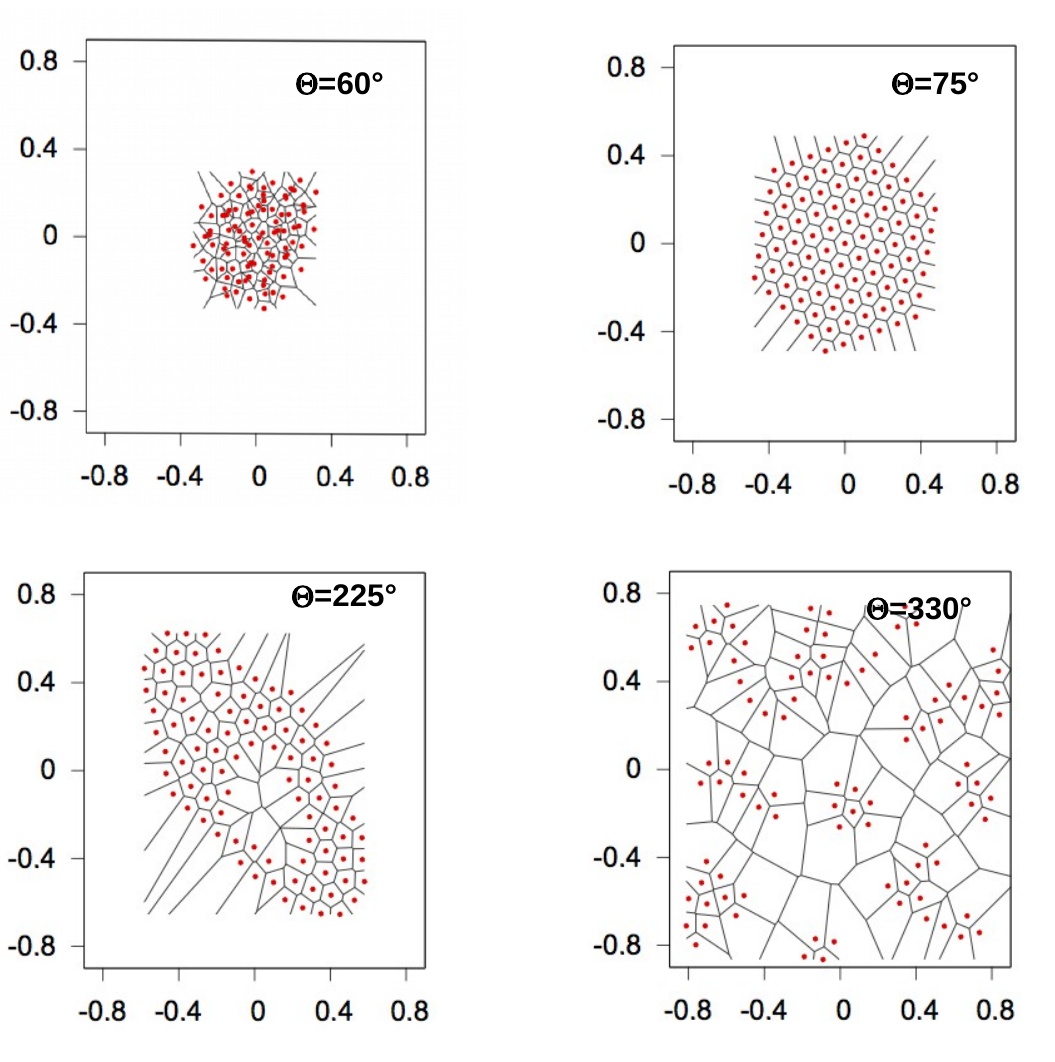}
\end{center}
\caption{\small {Typical configurations of the stationary states for different $\Theta$ values. 
The red dots represent individuals' positions, the lines the corresponding Voronoi tessellation.
}
}
\label{fig_final}
\end{figure}

Figure~\ref{fig_Dynamics} shows the time evolution of these order parameters
for two typical $\Theta$ values.
For $\Theta=75^{\circ}$, after a rapid transient,  the system 
enters in a state characterised by some level of cohesion and order.
This is shown by higher values of $\psi_6$ and $N$.
Even if in the case of $\psi_6$ the parameter value is very noise, we can distinct
a plateau which corresponds to higher level of order.
Looking at the parameter $C$ we can observe a
clear plateau with a reduction in the 
area occupied by individuals which reaches 80\%.
A high 
cohesion of the swarm is rapidly obtained and maintained.
At very long times, the system reaches an absorbing state
where all the individuals' distances with all their topological neighbours are equal to $D$. 
The interaction arranges the swarm in the densest way,
a perfect sixfold ordered structure (see Figure~\ref{fig_final}).
This is reported by an abrupt jump of the parameters to the values $\psi_6=1$
and $N=5.27$.
In contrast, $C$ maintains a value comparable with the one reached after 
leaving the first transient period.

For $\Theta=330^{\circ}$ the parameters values do not change significantly.
In fact, neither ordering, nor a general group contraction is possible. 
As can be seen in Figure~\ref{fig_final}, a fragmented swarm appears, 
where the initial group splits 
in different clusters and cohesion is lost. 

The outlined deep differences
present in the system dynamics
in dependence of  the 
attention field angle $\Theta$
can be 
studied also looking at the 
final configurations  attained 
by the groups (see Figure~\ref{fig_final}).
This analysis was systematically carried on 
in a previous  paper \cite{edgardo}.
Those results can be summarised
highlighting four different possible final phases.
For $\Theta<60\,^{\circ}$ a disordered, highly dense, connected phase appears.
For $60\,^{\circ}<\Theta<220\,^{\circ}$ it changes to a crystal-like phase, 
with  hexagonal patterns.
A low ordered phase with the presence of 
holes and ruptures is attained for $220\,^{\circ}<\Theta<300\,^{\circ}$.
Finally, for $\Theta>300\,^{\circ}$, a fragmented
swarm emerges, where the initial group splits 
into different clusters and cohesion is lost. 

The transition between these different phases 
can be characterised by looking at the 
 $\psi_6$, $N$ and $C$ values.
In this work, we are interested in unfolding an in deep
analysis of the transition between the crystal-like phase
and the low ordered phase. 
At sufficiently long times, this shift is characterised by the passage from an 
absorbing state, characterised by a single densely packed cluster of individuals, 
to an active, but slow, stationary state of disordered and/or fragmented clusters.
This transition is particularly relevant from the biological point of view,
because it discriminates between attention field angles
which efficiently allow for a single densely packed cluster,
with a sensitive reduction of the domains of danger,
from the ones which cannot generate effective cohesion.
For this reason, we analyse the transition looking at 
the $C$ value, the order parameter which better 
characterises the level of cohesion.
$C$ is measured 
in the plateau region, which can be considered
as a stationary state.
We do not wait to reach the final
configuration. 
In principle, the final 
configurations of the system 
are more interesting form a theoretical point of view, 
as they allow a precise description of the 
ordering property of the model.
On the other hand, from an empirical point of view, the living configurations
present all along the plateau region are the ones that can
be directly compared with the behaviour of real swarms.
With this choice, we are also able to speed up our measures,
even if the simulation time continues to be very large.
In fact, the system presents very slow relaxation time close to the transition. 
For this reason, we analyse the state reached after running $10^6$ 
Monte Carlo steps. Moreover, the algorithm for identifying
the nearest neighbour in the attention field is costly.
Finally, for estimating the $\langle C \rangle$ value, we must realise ensemble 
averages, 
and not time averages.
Such slow numerical analysis forces us to adopt $P$ values 
limited between 91 and 721.
Besides the $\langle C \rangle$ values, we estimate their fluctuations 
$\chi$, 
which are obtained 
by evaluating $\chi = L^{2}(\langle C^{2} \rangle - \langle C \rangle^{2})$.

\begin{figure}[h]
\begin{center}
\includegraphics[width=0.75\textwidth, angle=0]{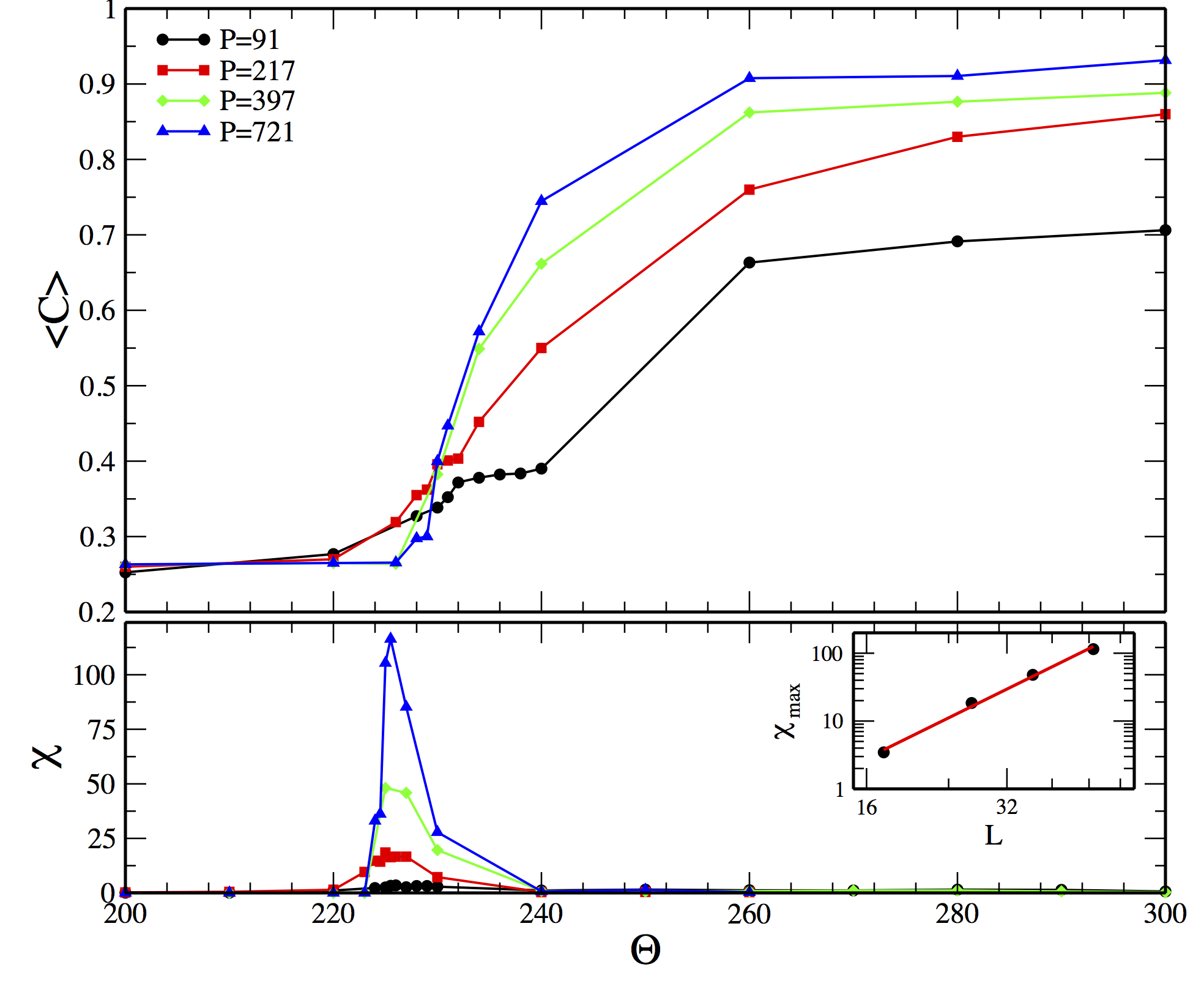}
\end{center}
\caption{\small 
Top: $\langle C \rangle$ in dependence of $\Theta$ for different system sizes.
Bottom: The fluctuation $\chi$ in dependence of $\Theta$.  In the inset, 
logarithmic plot of 
the scaling of the maxima of $\chi$ in dependence of $L$. The fluctuation peaks grow following a power-law (continuous line).
We remember that the density is fixed to 0.3, which implies that $L=\sqrt{P/0.3}$.
Each point is averaged over 100 simulations.
}
\label{fig_transition}
\end{figure}

In Figure~\ref{fig_transition} we exhibit the behaviour of the order parameter 
and its fluctuations as a function of $\Theta$ for different system sizes.
One can observe the typical behaviour of a phase transition: the order parameter presents
an evident shift near a critical value $\Theta_{c}$, which corresponds to a clear
peak in the fluctuation values. 
These peaks $\chi_{max}$ grow following a power-law  for increasing values of $L$, 
a behaviour typical of a classical phase transitions \cite{privman} (see Figure~\ref{fig_transition}).
Finally, as in a genuine phase transition, the curves exhibit 
the characteristic steepest aspect of the order parameter shift 
when the system size is increased.
In contrast, the expected dependence of the critical value
of $\Theta$  on the system size is not clearly 
identifiable.
This is probably due to the fact that 
its dependence is on a sharp scale of angles,
which we are not able to identify with 
the discretisation of our points and for the 
relative small difference among the considered system sizes.

\begin{figure}[h]
\begin{center}
\includegraphics[width=0.75\textwidth, angle=0]{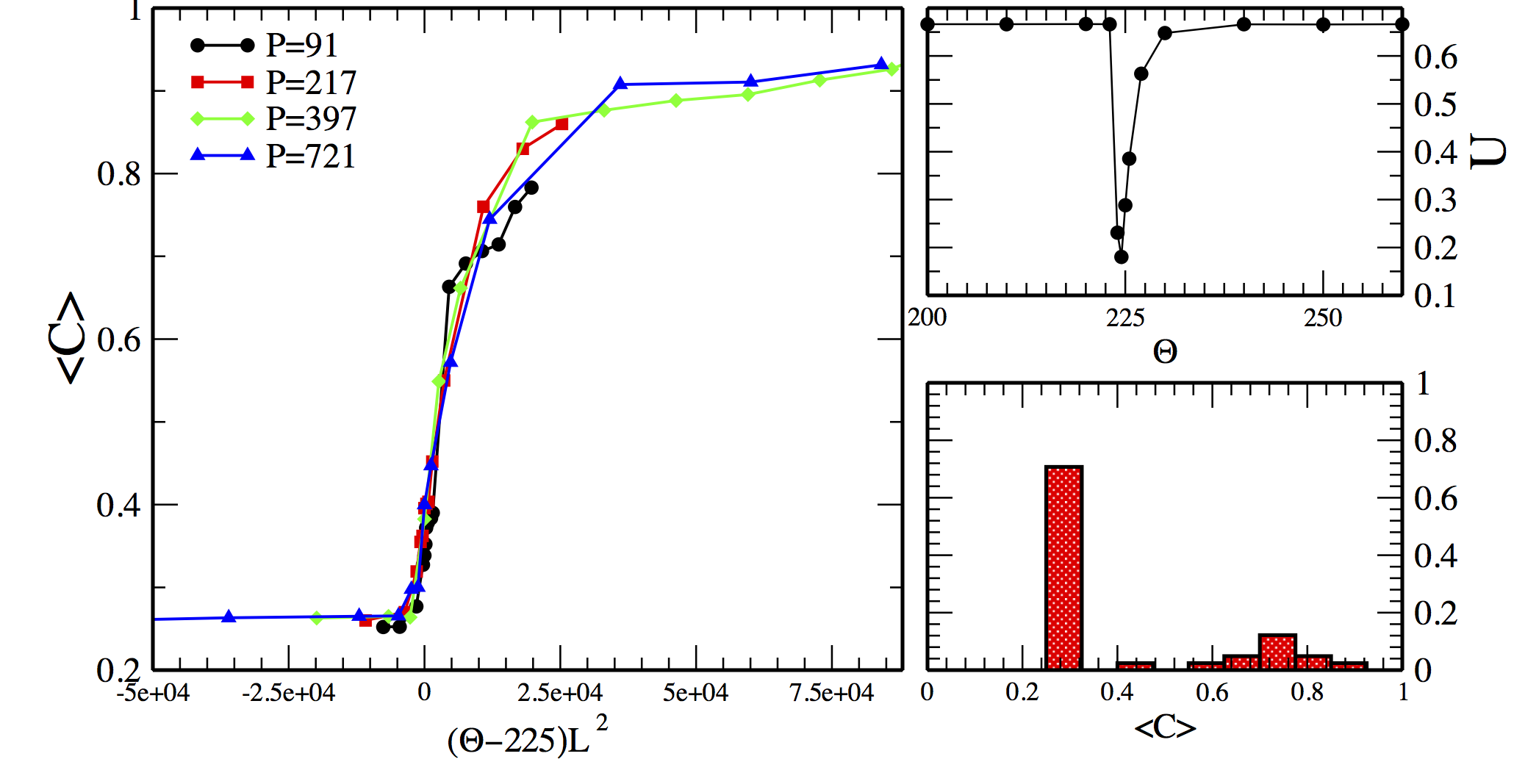}
\end{center}
\caption{\small 
Left: Rescaled plot for $\langle C \rangle$.
Top right: Binder cumulant as a function of $\Theta$ for P=721 
showing the characteristic well-defined minimum near the transition. 
Results are averaged over $100$ samples.
Bottom right: a typical example of the shape of the probability distribution function of 
$C$ near the minimum of the Binder cumulant.
}
\label{fig_scaling}
\end{figure}

We try to characterise the transition order 
looking at the Binder 
cumulant $U =1 - \frac{\langle C^{4}\rangle }{3\langle C^{2}\rangle ^{2}}$.
As expected for discontinuous phase transitions \cite{binder}, 
the Binder cumulant exhibits a sharp drop toward negative
values (see Figure~\ref{fig_scaling}). 
This minimum 
is generated by the coexistence of two phases
near 
$\Theta_{c}$.

Following this indication, we verify if the order parameter $\langle C \rangle$ presents the typical scaling of a discontinuous transition near the transition point, a standard procedure for equilibrium finite-size scaling analysis \cite{binder}. 
The scaling plot should be obtained by simply considering the rescaled control
 parameter $(\Theta-\Theta_{c})L^d$, where $d$ is the system dimension.
As shown in Figure~\ref{fig_scaling}, it is possible to obtain a reasonable collapse which satisfies this relation. In fact, data roughly collapse to a single curve, strongly suggesting the validity of the finite-size scaling ansatz expected for a discontinuous transition.
In the case of discontinuous phase transitions in equilibrium systems,  
near the transition point, 
fluctuation peaks should present power-law scaling  of the form: 
$\chi_{max} \sim  L^d$. 
We verified the power-law scaling of the fluctuation peaks,
but, for our out-of-equilibrium system, the exponent is 
intriguing close to $3.4$ 
(see the inset in Figure~\ref{fig_transition}).


The hallmark of a collective process can be characterised
measuring 
correlations.
In fact, we expect that, for generating a collective response,  
individuals must be able to influence their interaction partners 
to drive behavioural changes on a well defined 
spatial scale.
This measure is particularly relevant for our model
where it is not easy to perceive a clear directional order
in individuals movements, as we are describing only relative movements, 
and the gazing directions
are taken randomly. Despite these facts, 
a clear strong correlation can be detected.

To accomplish this, we measure correlation in the directions of individuals' motions.
The  correlation function 
measures to what extent the motion of individual $i$ is correlated 
to that of $j$. We used a definition inspired by the study of Attanasi {\it et al.} \cite{attanasi}:
\begin{equation}
F(r)=\frac{\sum_{i\ne j}^P \delta(r-r_{ij}) \;\; \delta v_i\cdot \delta v_j }{\sum_{i\ne j}^P \delta(r-r_{ij})}
\end{equation}
where $\delta v_i=\frac{{\bf x}^t_i-{\bf x}^{t+\Delta t}_i}{\Delta t}-{\bf V}$, and {\bf V} is the overall velocity of the group; 
$\delta(r-r_{ij})$ is the Kronecker's delta, which is equal to 1 if $r<r_{ij}<r+dr$
and zero otherwise, and $dr$ is the selected space discretisation unit.

\begin{figure}[]
\begin{center}
\includegraphics[width=0.5\textwidth, angle=0]{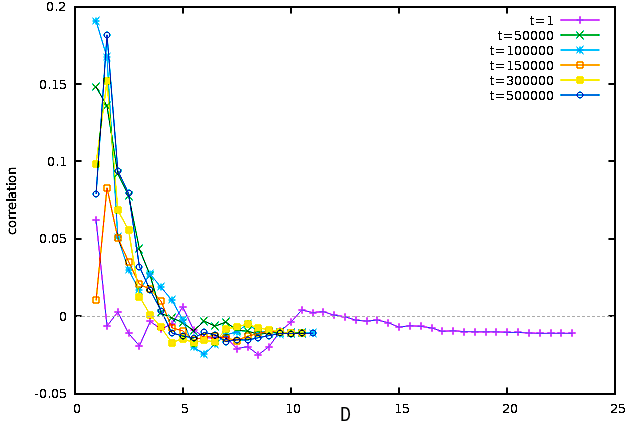}
\end{center}
\caption{\small The  correlation function measured at different
simulational time steps with the distance expressed in unit of $D$.
Note, after the
transient, the reduction in the group size. 
P=721, 
$\Theta=225^{\circ}$.
}
\label{fig_correlation}
\end{figure}

The behaviour of $F(r)$ is reported in Figure~\ref{fig_correlation},
measured at different times of the simulation.
We can observe how the correlation is practically null for
very small time steps, and it grows, for higher times, towards strong 
positive values at short distances. 
The correlation crosses zero at a distance equal to five $D$.
This means that neighbours inside a circle of radius $5D$,
tend to align their directions of motion.
As we are analysing a simulation near $\Theta_c$, in general,
we expect that individuals have a mean nearest neighbour distance
close to $D$. This fact suggests that the influence between individuals' 
directions of motion goes far beyond the first neighbour,
even if the model interaction operates only with the first topological neighbour.
For distances greater than five, $F(r)$ relaxes to very small negative 
values, which describe a general tendency to compact the group,
manifested by a direction of motion tending to tighten individuals' distances.\\
 
In summary, we have presented a numerical study of a simple model,
which describes collective movement in alarmed animals groups,
with the intent of clearly characterising the nature of the transition displayed 
by this system.
Our analysis shows that the model effectively presents a non-equilibrium phase transition between  a perfectly regular absorbing state and a low ordered active one.
In the first phase 
swarms are able to attain  a single densely packed cluster of individuals which can well describe the behaviour of some group of alarmed animals. 
The analysis is carried out looking at a control parameter which measures
the degree of cohesion
of the swarms.
A clear critical value of the angle which controls the attention field 
is identified. This value is consistent with a previous qualitative analysis 
realised looking at other order parameters \cite{edgardo}. 
The nature of the phase transition has been studied varying the system size.
The fluctuations of the order parameter show a power-law behaviour.
The form of the Binder cumulant, the coexistence of different phases 
near the critical point and the collapse of the scaling plot of 
$\langle C \rangle$ suggest that the system undergoes a 
discontinuous transition. However, the unexpected 
behaviour of the fluctuations scaling behaviour suggests prudence.

Finally, we 
described the correlations in the directions of individuals' motions, 
which show strong positive values at short distances. 
Hence correlations are relevant enough to prove that the individuals' effective perception range 
goes far beyond the model interaction range.
Despite our model implements random gazing directions, 
a clear strong influence between individuals can be measured.
All these  elements suggest that 
a stimulus generated by an environmental perturbation
can be perceived at a collective level, 
causing a general change in the group
behaviour. 
These results are not only relevant for this specific model but also, 
they show, more generally, how a model purely based on a repulsion/attraction positional force
can generate correlations analogous to the ones measured in field 
observations \cite{attanasi,attanasi2}.
This is an important result, which shows the relevance of positional interaction
models for describing swarms collective movements
and it sheds a new light on the possible 
nature of the interaction present in such systems.



\end{document}